\documentclass{article}
\usepackage{spconf,amsmath,graphicx}
\usepackage{booktabs}
\usepackage{hyperref}
\usepackage[hyphenbreaks]{breakurl}
\usepackage{pgfplots}
\usepackage{tikz}
\usepackage{subcaption}
\usepackage{ragged2e}
\usepackage{multirow}

\newcommand{\etal}{\emph{et al.}\ }

\pgfplotsset{compat=newest}
\usetikzlibrary{
    patterns,
    shapes.geometric
}
\usepgfplotslibrary{groupplots}

\title{Toward Degradation-Robust Voice Conversion}
\name{Chien-yu Huang$^{*1}$,
      Kai-Wei Chang$^{*2}$,
      Hung-yi Lee$^{12}$
      \thanks{$^*$ These authors contributed equally.}}
\address{$^1$Department of Electrical Engineering, National Taiwan University, Taiwan\\
$^2$Graduate Institute of Communication Engineering, National Taiwan University, Taiwan}

\begin{document}
\ninept
\maketitle
\begin{abstract}
Any-to-any voice conversion technologies convert the vocal timbre of an utterance to any speaker even unseen during training.
Although there have been several state-of-the-art any-to-any voice conversion models, they were all based on clean utterances to convert successfully.
However, in real-world scenarios, it is difficult to collect clean utterances of a speaker, and they are usually degraded by noises or reverberations.
It thus becomes highly desired to understand how these degradations affect voice conversion and build a degradation-robust model.
We report in this paper the first comprehensive study on the degradation robustness of any-to-any voice conversion.
We show that the performance of state-of-the-art models nowadays was severely hampered given degraded utterances.
To this end, we then propose speech enhancement concatenation and denoising training to improve the robustness.
In addition to common degradations, we also consider adversarial noises, which alter the model output significantly yet are human-imperceptible.
It was shown that both concatenations with off-the-shelf speech enhancement models and denoising training on voice conversion models could improve the robustness, while each of them had pros and cons.
\end{abstract}
\section{Introduction}
\label{sec:intro}

Voice conversion (VC) alters the vocal timbre of an utterance while preserving its linguistic content or phoneme structure.
These technologies were boosted with the help of deep learning in recent years \cite{chou2019one, qian2019autovc, lin2021fragmentvc, lin2021s2vc}.
Among them, any-to-any VC aims to convert the vocal timbre of an utterance to another speaker not necessarily seen during training, which is thus more challenging but closer to real applications.
The recent success of any-to-any VC heavily based on clean utterances, which are not distorted by noises or other degradations.
However, it is difficult to obtain clean utterances without a proper environment or professional recording equipment in real-world scenarios.
The converted utterance can be severely distorted in both linguistic content and vocal timbre due to the degraded inputs.
There may be mispronounced phonemes, and humans may not understand the content at all in the worst case.
Besides, the converted utterance may not sound like being produced by the target speaker, which means a failed VC.
All these defects constrained the applications of VC technologies.

In this paper, we study on and enhance the degradation robustness of any-to-any VC models.
We study how various degradations affect VC in linguistic content and vocal timbre.
We then propose two approaches to strengthening the robustness of VC models.
The first one is the concatenation with speech enhancement (SE).
We utilize off-the-shelf SE models as a preprocessing step to remove several degradations in utterances and thus improve the VC quality.
The second one is denoising training, which aims to enhance the robustness without modifying the network architecture or introducing additional parameters.
Specifically, we randomly apply different degraded data augmentation during training to explicitly force the model to filter the degradations and perform VC correctly.

We conducted both objective and subjective evaluations on two state-of-the-art any-to-any VC models with degraded inputs.
The evaluation results showed that the two proposed approaches help VC models perform significantly better across various degradations.
Interestingly, it was found that metrics used in SE did not necessarily correlate to their benefits to VC.
Meanwhile, the proposed approaches improve the degradation robustness and enhance the conversion quality with clean inputs.

We also studied adversarial noises in this paper. 
Different from common degradations, adversarial noises are human-imperceptible yet much powerful to alter the model output drastically.
It was shown that VC models fail to properly convert the vocal timbre when these noises are added to inputs \cite{huang2021defending}.
To defend VC models from such attacks, we adopted the two proposed approaches and further combined adversarial and denoising training.
Through the experiments, we found that SE models and denoising training could directly eliminate adversarial noises well.
In contrast, adversarial training increased training time yet did not show apparent improvements.

\section{Related Works}
\label{sec:related-works}

Conventional VC needs parallel data for training.
However, such data is difficult to collect as the number of speakers grows.
To this end, Chou \etal \cite{chou2018multi} adopted domain adversarial training to obtain disentangled representations of linguistic content and speaker.
CycleGAN-VC \cite{kaneko2018cyclegan} used GAN and cycle-consistency to obtain high-quality and linguistically meaningful outputs.
StarGAN-VC \cite{kameoka2018stargan} then further performed many-to-many VC with conditional input.
However, all these are limited to speakers seen during training, and their applications are thus constrained.

On the other hand, any-to-any VC converts between arbitrary speakers even unseen during training.
AdaIN-VC \cite{chou2019one} manipulated vocal timbre with adaptive instance normalization.
\textsc{AutoVC} \cite{qian2019autovc} used a pre-trained speaker encoder and an information bottleneck to encode inputs properly.
FragmentVC \cite{lin2021fragmentvc} leveraged Transformer architecture to build exemplar-based VC, and S2VC \cite{lin2021s2vc} further improved it with self-supervised representations, achieving state-of-the-art performance.

Noise-robust VC has been studied before neural networks were widely adopted \cite{aihara2015noise}, but there are still few studies on any-to-any VC.
Huang \etal \cite{huang2021far} studied the robustness of several recent VC models in various perspectives.
Nevertheless, they only conducted preliminary experiments with a few types of artificial noise, not real-world noises and other degradations as in this paper.
On the other hand, Voicy \cite{mottini2021voicy} overcame noisy reverberant conditions by introducing phonetic and acoustic-ASR encoders as additional modules, where transcriptions were required, however.
Besides, none of these previous works has investigated how to enhance the adversarial robustness, which also plays an important role in VC \cite{huang2021defending}.

\section{Degradation-robust Voice Conversion}
\label{sec:deg-robust-vc}

The general framework of any-to-any VC models follows encoder-decoder architecture \cite{chou2019one, qian2019autovc, lin2021fragmentvc, lin2021s2vc}.
The content encoder encodes linguistic content from a source utterance produced by the source speaker, while the speaker encoder extracts vocal timbre from a target utterance produced by the target speaker.
The decoder then receives encoder outputs and generates a converted utterance that contains the same linguistic content as the source utterance but sounds like being produced by the target speaker.

Empirically, a successful conversion relies on clean utterances for both source and target.
However, in real-world scenarios, utterances are inevitably degraded.
In this case, the converted utterance differs from the expected one, and its quality dropped significantly in both linguistic content and vocal timbre.

\subsection{Speech enhancement concatenation}
\label{ssec:se-concat}

To reduce the impact of degradations, the most straightforward way is to adopt an off-the-shelf SE model, which removes them from utterances.
We first process inputs with an SE model to obtain enhanced utterances that contain fewer degradations.
The VC model then receives these enhanced utterances, and hopefully, the enhanced utterance would be well converted.

However, the SE model only learns to remove the noise, so there is no guarantee that its output would be suitable for VC.
On the other hand, the VC model has never seen enhanced utterances during the training. 
The SE model may remove some critical speech characteristics for VC, and thus the enhanced utterances hamper VC results.
Moreover, adopting an SE model introduces much more parameters and makes the whole system larger, and thus is unfriendly for deployment on mobile devices in real-world scenarios.

\subsection{End-to-end denoising training}
\label{ssec:e2e-dnt}

We then propose end-to-end denoising training that reinforces VC models by adopting new training objectives and data augmentations to avoid the above concerns.
During the training, we randomly adopt data augmentation to generate degraded utterances.
Then, the VC model randomly receives clean or degraded utterances as input, and we calculate the loss with the authentic clean data in both cases.

Since reconstruction loss is the main objective for training most any-to-any VC models nowadays, our end-to-end denoising training makes the VC model like a denoising autoencoder, which receives degraded data as input and needs to reconstruct original clean data.
From another perspective, since the model recovers clean utterances from degraded ones, we can consider this approach a combination of training SE and VC in a single model.

\section{Adversarial Attack and Defense}
\label{sec:attack-defense}

We can make a VC model fail to convert vocal timbre properly by adding particular noises obtained via adversarial attack to target utterances \cite{huang2021defending}.
However, it is not yet investigated how to defend VC models against these attacks.
These noises are human-imperceptible yet able to alter the model output drastically, and we thus consider them also a particular degradation.
There are two scenarios in adversarial attack: untargeted and targeted attack.
An untargeted attack simply aims to alter the vocal timbre of converted utterances to make them not sound like being produced by the target speaker.
In contrast, a targeted attack further attempts to make the converted utterance sound like being produced by a specific third speaker other than the target speaker.

On the other hand, there are two scenarios in defense against adversarial attacks.
The first is passive defense, where we attempt to remove adversarial noises by additional data preprocessing while keeping the model unchanged, and the SE concatenation belongs to this scenario.
The second is proactive defense, which enhances the robustness from the training stage.
The most popular one is adversarial training, where the model is directly trained on adversarial examples \cite{goodfellow2015explaining}. 
Here we combine denoising training and adversarial training by considering adversarial noises an another augmentation.

\section{Experimental Settings}
\label{sec:exp-settings}

\subsection{Models}
\label{ssec:models}

We evaluated the proposed approaches on two state-of-the-art any-to-any VC models: AdaIN-VC\footnote{\url{https://github.com/cyhuang-tw/AdaIN-VC}} and S2VC\footnote{\url{https://github.com/howard1337/S2VC}}.
S2VC includes several self-supervised representations, and here we adopted CPC \cite{oord2018representation} version since it was reported to perform the best.
For SE, we chose off-the-shelf models pre-trained on different datasets: \textsc{Demucs} \cite{defossez2020real}, on Valentini \cite{valentini2017noisy} and DNS \cite{reddy2020interspeech}; MetricGAN+ \cite{fu2021metricgan}, on VoiceBank-DEMAND \cite{valentini2016investigating}; and Conv-TasNet \cite{luo2019conv}, on LibriMix \cite{cosentino2020librimix}.

The above VC models do not generate converted utterances in waveform directly.
To convert acoustic features to waveforms, we used a WaveRNN-based universal vocoder\footnote{\url{https://github.com/yistLin/universal-vocoder}} \cite{trueba2019towards} trained on LibriTTS \cite{zen2019libritts} (\emph{train-clean-100}), LJ Speech \cite{ljspeech17} and CMU Arctic \cite{kominek2004cmu} for both VC models.

\subsection{Degraded data augmentation}
\label{ssec:deg-data-aug}

We considered three common degradations: additive noise, reverberation, and band rejection implemented in WavAugment\footnote{\url{https://github.com/facebookresearch/WavAugment}}.
The augmentation procedure during the training was as follows.
We randomly augmented 60\% of the utterances in a mini-batch and remained the other 40\% intact.
For additive noise, we used DEMAND \cite{thiemann2013demand} dataset for various types of background noises.
For each utterance to be degraded, we first added noise with SNR uniformly sampled from 0, 5, 10, and 15.
Then, we applied reverberation (\emph{room-scale} from 0 to 100) and band rejection (bandwidth from 50 to 150, with the lower frequency, ranged from 100 Hz to 500 Hz) with the probability being 0.5 for each.

\subsection{Adversarial training}
\label{ssec:adv-training}

In the case adversarial training was combined with denoising training, we randomly applied embedding attack \cite{huang2021defending} to each utterance in a mini-batch, no matter already augmented, with probability 0.5.
However, the typical adversarial training is inefficient and time-consuming.
We thus resorted to \textit{fast} adversarial training \cite{wong2019fast} to speed up the training.
For simplicity, we fixed the step size $\alpha$ and the maximum amplitude of noises $\epsilon$ to be 0.001 and 0.005, respectively.

\subsection{Test scenarios}
\label{ssec:test-scenarios}

We selected 20 speakers from CSTR VCTK Corpus \cite{veaux2016vctk} and denoted it as VCTK-test.
These speakers were all unseen during the training, which is important to the performance evaluation of any-to-any VC.
The degraded data generation followed the same procedure described in Sec.~\ref{ssec:deg-data-aug} except for a different noise dataset, WHAM! \cite{wichern19_interspeech}, and SNR being 2.5, 7.5, 12.5, or 17.5 to ensure that both noise types and SNR values were unseen in denoising training.
We randomly sampled 250 conversion pairs (source and target) and performed VC in all objective evaluations.
As for subjective evaluation, we further sampled 25 utterances from the above 250 converted utterances, and each utterance was scored by five subjects individually.

We leveraged embedding attacks to test adversarial robustness.
For each conversion pair, we randomly sampled the third speaker from VCTK-test and performed a targeted attack, aiming to make the converted utterance sound like being produced by that speaker but not the target speaker.
We assumed the attacker had full access to architectures and parameters of VC models but was unaware of SE models, and thus all attacks were against VC models only.

\subsection{Evaluation metrics}
\label{ssec:evaluation-metrics}

An automatic speech recognition (ASR) system\footnote{\burl{https://huggingface.co/facebook/wav2vec2-large-960h-lv60-self}} was used to evaluate the naturalness of converted utterances.
Intuitively, suppose a converted utterance well preserved the linguistic content from the source utterance.
In that case, the character error rate (CER) from ASR should be low or very close to that of the source utterance.
Conversely, a higher CER indicates that the converted utterance was severely distorted and did not sound natural.

On the other hand, we adopted a speaker verification (SV) system\footnote{\url{https://github.com/resemble-ai/Resemblyzer}} to measure the vocal timbre of converted utterances.
It took a converted utterance and a target utterance as input and then generated two fixed-dimensional embeddings.
If the cosine similarity between the embeddings exceeded a pre-defined threshold, the two utterances were considered produced by the same speaker.
In this case, the conversion was considered successful.
The threshold was obtained by computing the equal error rate (EER) on the entire VCTK dataset.
We then defined speaker verification accept rate (SVAR) to be the percentage of successful conversions.

For subjective evaluation, we conducted two Mean Opinion Score (MOS) tests for naturalness and speaker similarity, respectively.
For naturalness, subjects were given either a converted utterance or an authentic utterance, and they determined the perceptual quality ranging from 1 to 5.
A high score means an utterance sounded natural.
As for speaker similarity, subjects listened to a target utterance and the converted utterance, and they answered whether the same speaker produced the two with a score from 1 (absolutely different) to 5 (absolutely same).

\section{Results}
\label{sec:results}

\subsection{Objective evaluation}
\label{ssec:obj-eval}

Table~\ref{tab:proposed-performance} shows the VC performances under different scenarios\footnote{S2VC used CPC representations if not specified.}.
The results of baseline VC models are also included for comparison.
In terms of baseline models, with degraded inputs, CER soared while SVAR dropped significantly, indicating that both linguistic content and vocal timbre of the converted utterances were severely altered and distorted.
We further included S2VC with wav2vec 2.0 \cite{baevski2020wav2vec} representations for comparison.
The performance drop on S2VC indicates that self-supervised representations, CPC and wav2vec 2.0 in this case, were not robust to degradations in terms of VC applications.
Furthermore, the performances of these representations on degraded data are correlated to those on clean data, showing that the choice of self-supervised representations is an essential factor in degradation robustness.

In SE concatenation, results were much different across SE models.
As a reference, Table~\ref{tab:se-performance} lists the performance of SE models on degraded VCTK-test on SE metrics.
The higher score in these metrics indicates better performance in terms of SE.
Among the three SE models, \textsc{Demucs} performed the best.
It effectively improved VC performance, as both CER and SVAR were close to those from clean inputs.
Conv-TasNet slightly fell behind but still helped VC to some extent.
Surprisingly, although MetricGAN+ obtained a great score on PESQ, it yielded a slight improvement in VC (especially SVAR).
This suggests that SE evaluation metrics do not always correlate to improvements on VC.
Then, we see that denoising training still worked well on both AdaIN-VC and S2VC with the performance on par with SE concatenation.
In addition, Table~\ref{tab:s2vc-clean} lists the performance of proposed approaches on clean data.
Interestingly, denoising training improved the CER much, while \textsc{Demucs} slightly improved SVAR.
Overall, both approaches improved VC performance well with different merits: SE concatenation did not involve training, while denoising training did not need additional parameters.

\begin{table}[ht]
    \footnotesize
    \centering
    \caption{The performance of SE concatenation and denoising training on degraded VCTK-test. Baselines are included for comparison.}
    \label{tab:proposed-performance}
    \begin{subtable}{\linewidth}
        \centering
        \caption{character error rate}
        \label{tab:cer}
        \begin{tabular}{ccccc}
            \toprule
            \multicolumn{2}{c}{\textbf{Scenarios}} & AdaIN-VC & S2VC & S2VC-W2V \\
            \midrule
            Clean & baseline      & 37.31\%          & 26.79\%          & 14.49\% \\
            \midrule
            \multirow{5}{*}{Degraded} & baseline   & 62.42\%          & 68.26\%          & 68.24\% \\
            {} & \textsc{Demucs}    & \textbf{43.83\%} & 35.71\%          & 24.46\% \\
            {} & MetricGAN+         & 53.56\%          & 47.84\%          & 36.69\% \\
            {} & Conv-TasNet        & 48.48\%          & \textbf{33.93\%} & \textbf{23.25\%} \\
            {} & Denoising & 45.41\%          & 38.94\%          & 25.59\% \\
            \bottomrule
        \end{tabular}
    \end{subtable}
    \justify
    \begin{subtable}{\linewidth}
        \centering
        \caption{speaker verification accept rate}
        \label{tab:svar}
        \begin{tabular}{ccccc}
            \toprule
            \multicolumn{2}{c}{\textbf{Scenarios}} & AdaIN-VC & S2VC & S2VC-W2V \\
            \midrule
            Clean & baseline      & 85.2\%          & 86.8\%          & 65.6\% \\
            \midrule
            \multirow{5}{*}{Degraded} & baseline   & 5.2\%           & 4.4\%           & 1.6\%\\
            {} & \textsc{Demucs}    & 64.4\%          & \textbf{77.2\%} & \textbf{59.6\%} \\
            {} & MetricGAN+         & 14.8\%          & 16.4\%          & 33.6\% \\
            {} & Conv-TasNet        & 58.0\%          & 71.2\%          & 55.6\% \\
            {} & Denoising & \textbf{66.0\%} & 71.6\%          & 53.2\% \\
            \bottomrule
        \end{tabular}
    \end{subtable}
\end{table}

\begin{table}[htb]
    \centering
    \caption{The evaluation results of SE models on degraded VCTK-test (additive noise only / all degradations).}
    \label{tab:se-performance}
    \begin{tabular}{cccc}
        \toprule
        \textbf{Metrics} & \textsc{Demucs} & MetricGAN+ & Conv-TasNet \\
        \midrule
        PESQ & \textbf{2.62} / \textbf{2.26} & 2.52 / 2.15 & 2.53 / 1.96 \\
        STOI & \textbf{0.96} / \textbf{0.95} & 0.90 / 0.91 & \textbf{0.96} / 0.94 \\
        SI-SDR (dB) & 16.67 / \textbf{9.99} & 2.79 / 2.47 & \textbf{17.5} / 9.45 \\
        \bottomrule
    \end{tabular}
\end{table}

\begin{table}[ht]
    \setlength{\tabcolsep}{3pt}
    \centering
    \caption{The performance of proposed approaches with S2VC on clean VCTK-test.}
    \label{tab:s2vc-clean}
    \resizebox{\linewidth}{!}{
    \begin{tabular}{cccccc}
        \toprule
        \textbf{Metrics} & baseline & \textsc{Demucs} & MetricGAN+ & Conv-TasNet & Denoising \\
        \midrule
        CER & 26.79\% & 26.95\% & 26.80\% & 27.35\% & \textbf{20.05\%} \\
        SVAR & 86.8\% & 88.4\% & 81.2\% & \textbf{89.2\%} & 83.6\% \\
        \bottomrule
    \end{tabular}
    }
\end{table}

\subsection{Subjective evaluation}
\label{ssec:sub-eval}

We further compared the two best results obtained previously by MOS tests: \textsc{Demucs} and denoising training on S2VC.
Table~\ref{tab:mos} lists the two MOS results of converted utterances using clean or degraded inputs.
Obviously, the original S2VC was not robust.
It performed much worse with degraded inputs, and these converted utterances sounded neither natural nor like being produced by the target speakers, leading to failed VC.
On the other hand, both \textsc{Demucs} and denoising training achieved much better results on both tests, which were comparable to those of original S2VC using clean inputs.
Specifically, SE concatenation performed better in naturalness, while denoising training obtained higher similarity scores.
Last, we see similar trends in objective and subjective evaluation results, strengthening the correctness of objective evaluations above.
The demo page and the source code are available at \url{https://cyhuang-tw.github.io/robust-vc-demo}.

\begin{table}[ht]
    \setlength{\tabcolsep}{2pt}
    \centering
    \caption{The MOS of S2VC along with different approaches.}
    \label{tab:mos}
    \begin{tabular}{cccccc}
        \toprule
        \multirow{2}{*}{\textbf{MOS}} & \multicolumn{2}{c}{Clean} & \multicolumn{3}{c}{Degraded} \\
        \cmidrule(lr){2-3}
        \cmidrule(lr){4-6}
        {} & Authentic & baseline & baseline & \textsc{Demucs} & Denoising \\
        \midrule
        Nat. & 4.71$\pm$0.05 & 3.54$\pm$0.08 & 1.45$\pm$0.07 & \textbf{3.32$\pm$0.09}& 3.26$\pm$0.09 \\
        Sim. & - & 3.67$\pm$0.10 & 2.74$\pm$0.07 & 3.38$\pm$0.11& \textbf{3.50$\pm$0.11} \\
        \bottomrule
    \end{tabular}
    \justify
    Nat.: naturalness, Sim.: speaker similarity.
\end{table}

\subsection{Performance analysis}
\label{ssec:performance-analysis}

The above analysis was based on fully-degraded utterances (with three degradations).
Table~\ref{tab:ablation} shows the VC performance when we only applied one degradation to input utterances.
Additive noise distorted the converted utterances the most because it led to the highest CER and the lowest SVAR in all models.
Reverberation also severely hampered VC, but it impacted CER more than SVAR, which matches the fact that it is hard to understand the content of a reverbed utterance because consecutive phonemes interfere with each other.
On the other hand, even with reverberation, we may still easily recognize the speaker.
Last, band rejection had the most negligible impact on both metrics, and it affected more on SVAR than it did on CER.
This is probably because the filtered frequency band did not contain important phoneme information, while high-frequency components were keys to different vocal timbre.

\begin{table}[ht]
    \footnotesize
    \centering
    \caption{The performance of VC models when exactly one degradation was applied to input utterances.}
    \label{tab:ablation}
    \begin{subtable}{\linewidth}
    \centering
    \caption{character error rate}
    \label{tab:ablation-cer}
    \begin{tabular}{ccccc}
        \toprule
        \multirow{2}{*}{\textbf{Degradation}} & \multicolumn{2}{c}{Original} & \multicolumn{2}{c}{Denoising Training} \\
        \cmidrule(lr){2-3}
        \cmidrule(lr){4-5}
        {} & AdaIN-VC & S2VC & AdaIN-VC & S2VC \\
        \midrule
        None & 37.31\% & 26.79\% & 34.69\% & 20.52\% \\
        Additive noise & 56.52\% & 60.58\% & 42.60\% & 33.70\% \\
        Reverberation  & 52.73\% & 42.73\% & 38.50\% & 24.70\% \\
        Band rejection & 37.40\% & 27.66\% & 35.30\% & 19.86\% \\
        \bottomrule
    \end{tabular}
    \end{subtable}
    
    \justify
    
    \begin{subtable}{\linewidth}
    \centering
    \caption{speaker verification accept rate}
    \label{tab:ablation-svar}
    \begin{tabular}{ccccc}
        \toprule
        \multirow{2}{*}{\textbf{Degradation}} & \multicolumn{2}{c}{Original} & \multicolumn{2}{c}{Denoising Training} \\
        \cmidrule(lr){2-3}
        \cmidrule(lr){4-5}
        {} & AdaIN-VC & S2VC & AdaIN-VC & S2VC \\
        \midrule
        None & 85.2\% & 86.8\% & 82.4\% & 84.8\% \\
        Additve noise  & 10.0\% & 6.8 \% & 67.6\% & 72.8\% \\
        Reverberation  & 39.6\% & 39.6\% & 79.2\% & 78.8\% \\
        Band rejection & 74.0\% & 83.2\% & 80.8\% & 81.2\% \\
        \bottomrule
    \end{tabular}
    \end{subtable}
\end{table}

\subsection{Adversarial robustness}
\label{ssec:adv-robustness}

We also evaluated the adversarial robustness of VC models.
Table~\ref{tab:attack-results} lists the SVAR of VC models under adversarial attacks.
Lower SVAR indicates more failed conversions and thus more successful attacks.
We only measured SVAR since the adversarial attacks aimed to alter the vocal timbre of converted utterances instead of linguistic content.
We see adversarial attacks worked well on the original VC models since the SVAR dropped significantly and was much lower than their original ones.
As we applied SE concatenation, the SVAR increased for both models (fewer failed conversions), though not as high as it was, showing that SE models not only suppressed common degradations but eliminated subtle adversarial noises to some extent.
On the other hand, denoising training also enhanced the adversarial robustness with comparable effectiveness to SE concatenation.
We then found that when integrating adversarial training, the SVAR did not increase further.
This shows that denoising training was already sufficient to make the VC model robust against adversarial attacks to some extent though no attack was involved in training.

\begin{table}[ht]
    \centering
    \caption{The SVAR of proposed approaches under embedding attack ($\epsilon = 0.005$).}
    \label{tab:attack-results}
    \begin{tabular}{cccccc}
    \toprule
    \textbf{Models} & - & \textsc{Demucs} & DNT & DNT + Adv. \\
    \midrule
    AdaIN-VC & 45.2\% & \textbf{79.2\%} & 78.8\% & 76.8\% \\
    S2VC & 78.4\% & \textbf{86.4\%} & 80.0\% & 80.0\% \\
    \bottomrule
    \end{tabular}
    \justify
    DNT: denoising training,
    Adv.: adversarial training.
\end{table}

\section{Conclusions}
\label{sec:conclusions}

Although substantial improvements have recently been achieved in VC technologies, their applications are limited due to various degradations in real-world scenarios.
This paper presents the first comprehensive analysis of the degradation robustness of state-of-the-art any-to-any VC models.
Two approaches are proposed to enhance the model performance against different degradations.
The evaluation results showed that even though SE models had similar performance in terms of SE evaluation metrics, their effectiveness on VC might vary a lot.
On the other hand, we can significantly enhance the degradation robustness of VC models without modifying model architecture or expanding model size through denoising training.
Besides, the proposed approaches also enhanced the adversarial robustness effectively.

\section{Acknowledgement}
\label{sec:acknowledgement}

We thank National Center for High-performance Computing (NCHC) of National Applied Research
Laboratories (NARLabs) in Taiwan for providing computational and storage resources.

\bibliographystyle{IEEEbib}
\bibliography{refs}

\end{document}